\documentclass[conference]{IEEEtran}
\IEEEoverridecommandlockouts
\usepackage{cite}
\usepackage{amsmath,amssymb,amsfonts}
\usepackage{clrs+} 
\usepackage{graphicx}
\usepackage{textcomp}
\usepackage{algorithm}
\usepackage[noend]{algpseudocode}
\usepackage{enumitem}

\newcommand{\qc}{\c{c}}
 \usepackage{fancyhdr} 
 \usepackage{textcomp}

\usepackage{color}

\usepackage{color, soul} 
\soulregister\emph7
\soulregister\cite7
\soulregister\ref7

\def\BibTeX{{\rm B\kern-.05em{\sc i\kern-.025em b}\kern-.08em
    T\kern-.1667em\lower.7ex\hbox{E}\kern-.125emX}}
\begin{document}
\title{Extreme Scale De Novo Metagenome Assembly}

\author{
Evangelos Georganas$^{*}$, Rob Egan$^\dag$, Steven Hofmeyr$^\ddag$, Eugene Goltsman$^\dag$, Bill Arndt$^\P$,  Andrew Tritt$^\P$\\ Ayd\i n Bulu\qc$^{\ddag}$, Leonid Oliker$^\ddag$, Katherine Yelick$^{\ddag}$\vspace{5pt}\\
 \normalsize 
 {$^*$ Parallel Computing Lab, Intel Corp.}\vspace{1pt} \\
 {$^\ddag$Computational Research Division\ /\ $^\dag$Joint Genome Institute, Lawrence Berkeley National Laboratory, USA}\vspace{1pt}\\
 {$^\P$ National Energy Research Scientific Computing Center, USA}}

\maketitle

\begin{abstract}
Metagenome assembly is the process of transforming a set of short, overlapping, and potentially erroneous DNA segments from environmental samples into the accurate representation of the underlying microbiomes's genomes. State-of-the-art tools require big shared memory machines and cannot handle contemporary metagenome datasets that exceed Terabytes in size. In this paper, we introduce the MetaHipMer pipeline, a high-quality and high-performance metagenome assembler that employs an iterative de Bruijn graph approach. MetaHipMer leverages a specialized scaffolding algorithm that produces long scaffolds and accommodates the idiosyncrasies of metagenomes. MetaHipMer is end-to-end parallelized using the Unified Parallel C language and therefore can run seamlessly on shared and distributed-memory systems. Experimental results show that MetaHipMer matches or outperforms the state-of-the-art tools in terms of accuracy. Moreover, MetaHipMer scales efficiently to large concurrencies and is able to assemble previously intractable grand challenge metagenomes. We demonstrate the unprecedented capability of MetaHipMer by computing the first full assembly of the Twitchell Wetlands dataset, consisting of 7.5 billion reads -- size 2.6 TBytes.
\end{abstract}

\section{Introduction}
\label{sec:introduction}
Metagenomics is currently the leading technology in studying uncultured microbial diversity and
delineating the structure and function of the microbiome, which is the collection of microorganisms
in a particular environment, e.g. the body. Improvements in sequencing techonolgy (in terms of
cost reduction) have significantly outpaced Moore's Law~\cite{wetterstrand2013dna}, enabling the 
collection of large numbers of human and environmental samples that comprise hundreds or even
thousands of microbial genomes. Assembly of metagenome samples into long contiguous sequences is
critical for the identification of long biosynthetic clusters and gene finding in
general~\cite{donia2014systematic}, and is also key for enabling the discovery of novel lineages of life and viruses~\cite{eloe2016metagenomics}. However, for most microbial samples, there is no existing reference genome, so a first step in analysis is \emph{de novo} assembly: transforming a set of short, overlapping, and potentially erroneous DNA segments (called \emph{reads}) from these samples into the accurate representation of the underlying microbiomes's genomes.

The exact de novo assembly of a genome is an NP-hard problem in
general~\cite{kececioglu1995combinatorial}. The metagenome assembly is further complicated by
identical sequences across different genomes, polymorphisms within species and the variable
abundance of species within the sample. The bioinformatics community has therefore developed special
algorithms~\cite{Ray-Meta,peng2012idba,namiki2012metavelvet,haider2014omega,li2015megahit,nurk2016metaspades,SOAPdenovo2,MEGAHIT1.0}
to overcome these challenges. Nevertheless, the vast majority of these tools are not parallelized
for distributed-memory systems. As a result, they require specialized, large shared-memory machines
with hundreds of GB of memory in order to deal even with modestly-sized metagenomic datasets. At the
same time, the concurrency of these tools is limited by the core count of a single node (typically
10s of cores) and consequently the execution times even for small datasets are in the order of
days. The only exception among the metagenome assemblers is Ray Meta, which is a tool designed for
distributed-memory systems. However, Ray Meta is not scalable to massive concurrencies~\cite{hipmer}
and its assembly quality has been shown to be worse compared to other state-of-the-art
tools~\cite{nurk2016metaspades}. Currently, existing tools for high-quality metagenome assembly are
incapable of processing large, realistic datasets due to the large memory and computational requirements.

In this work we introduce MetaHipMer, the first massively
scalable, high quality metagenome assembly pipeline. MetaHipMer implements an iterative de Bruijn graph approach similar to IDBA-UD~\cite{peng2012idba} and Megahit~\cite{li2015megahit,MEGAHIT1.0} to generate long, contiguous and accurate sequences called \emph{contigs}. MetaHipMer also performs specialized scaffolding to stitch together multiple contigs and further increase contiguity. 
The result of our work is the first distributed-memory metagenome assembler that achieves comparable
quality to the state-of-the-art tools, but scales efficiently to tens of thousand of cores and
decreases the execution times by orders of magnitude compared to single-node tools. While our
primary novelty is in enabling the high-quality assembly of larger datasets that current tools
struggle to deal with, MetaHipMer can also be seamlessly executed on shared memory machines. Overall this study makes several contributions including:
\vspace{-2mm}
\begin{itemize}[labelindent=0em,labelsep=0.1cm,leftmargin=*]
\item An iterative contig generation algorithm that relies on efficient, distributed hash tables,
  and combines best practices from state-of-the-art
  tools with new ideas tailored for metagenome datasets. The new algorithm obviates the need for an
  expensive, explicit input error-correction step that other tools rely on. This iterative approach
  allows MetaHipMer to directly handle large metagenome samples without an expensive error correction step,
  which could eliminate some data that would be valuable in the assembly.
  
\item A new parallel graph algorithm (also using distributed hash tables) that operates on partially assembled
  data to resolve ambiguities and errors and further extend the assembled regions in a process
  called scaffolding. This new algorithm also optimizes the accurate assembly of highly conserved
  ribosomal regions, which are the cornerstone for various downstream metagenomic analyses.
  
\item End-to-end parallelization of the entire MetaHipMer pipeline. This result is enabled by efficient
  distributed hash tables and high-performance data structures, with a combination of algorithmic
  techniques from parallel computing, including work partitioning via connected components and load balancing, as well as
  the use of a Partitioned Global Address Space language called Unified Parallel C (UPC).
  
\item Unprecedented scalability results on NERSC's Cori supercomputer, a Cray XC40 system, using
  synthetic and real world datasets. This work also presents the first whole assembly of
  Twitchell Wetlands, a complex, massive-scale metagenome dataset consisting of 7.5 billion
  reads with size 2.6TB. The Wetlands assembly highlights the new capabilities that MetaHipMer enables
  for metagenome analyses.
  
\end{itemize}

%
%

\section{Iterative Contig Generation}
\label{sec:contigging}

\begin{algorithm}[!t]
\begin{algorithmic}[1]
\State{\textbf{Input:} A set of paired reads $R$}
\State{\textbf{Output:} A set of contigs $C$}
\State $C \gets \emptyset$
\State $\id{prev\_k\mbox{-}mer\_set} \gets \emptyset$
\For {$k=k_{min}$ {\textbf{to}} $k_{max}$ with step $s$}
\State $k\mbox{-}mer\_set \gets \ $\Call{K\mbox{-}merAnalysis}{$k, R$}
\State $\id{new\_k\mbox{-}mers} \gets $\Call{Merge}{$\id{k\mbox{-}mer\_set} , \id{prev\_k\mbox{-}mer\_set}$}
\State $C_k \gets \ $\Call{DeBruijnGraphTraversal}{$\id{new\_k\mbox{-}mers}$}
\State $C'_k \gets \ $\Call{BubbleMerging}{$C_k$}
\State $C''_k \gets \ $\Call{IterativeGraphPruning}{$C'_k$}
\State $\id{Alignments_k} \gets \ $\Call{AlignReadsToContigs}{$R,C''_k$}
\State $C \gets \ $\Call{LocalAssembly}{$R, C''_k, \id{Alignments_k}$}
\State $\id{prev\_k\mbox{-}mer\_set} \gets $\Call{ExtractKmers}{$C,k+s$}
\EndFor
\State $\Call{Return}{}\ C$
\end{algorithmic}
\caption{Iterative contig generation}
\label{alg:cont}
\end{algorithm}

\begin{figure}[t!]
\centering
\includegraphics[width=\columnwidth]{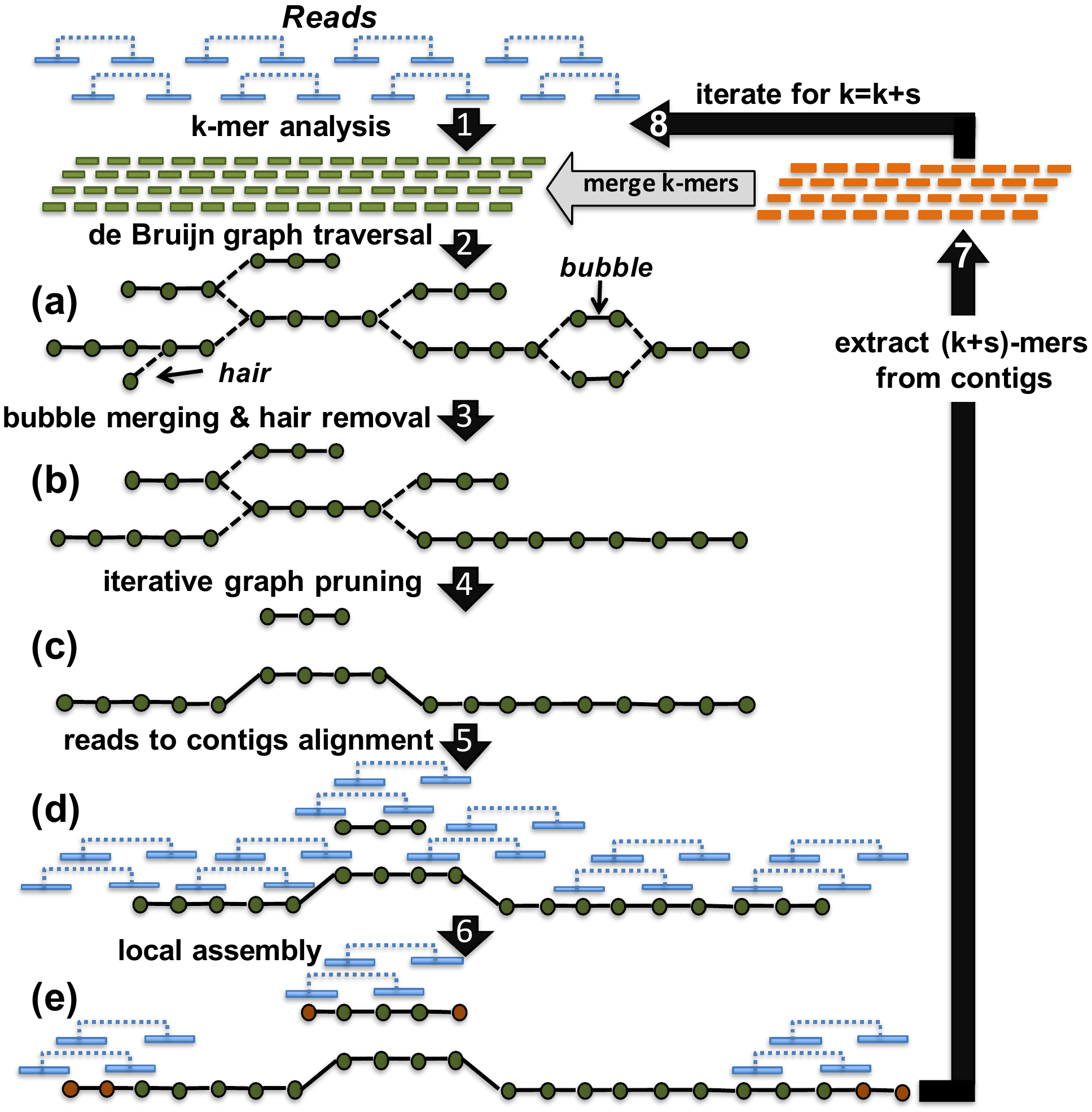}
\caption{Iterative contig generation workflow in MetaHipMer.}
\vspace{-10pt}
\label{fig:pipeline}
\end{figure}

Before diving into MetaHipMer's algorithm, we introduce some terminology that is used
throughout the paper. \textbf{Reads} are typically short fragments of DNA sequence that are produced
by DNA sequencers; current sequencing technology can only read the genome in fragments. These reads
contain errors and may also come in pairs (e.g.\ see Figure~\ref{fig:pipeline}, where pairs of reads
-- light blue pieces -- are connected with dashed lines). Reads are strings of four possible
nucleotides/bases: A, C, G and T. A read \textbf{library} is a source of DNA template fragments that
the reads are generated from and is typically characterized by an \textbf{insert size}, which is the
distance between the two ends of the paired reads. Every genomic region/sample is covered by
multiple, overlapping reads, which is necessary to identify and exclude errors from the reads. \textbf{$K$-mers} are short overlapping substrings of length $k$ that are typically extracted from reads. A \textbf{de Bruijn graph} is an efficient way to represent a sequence in terms of its $k$-mer components. In this type of graph, vertices are $k$-mers and two $k$-mers that overlap in $k-1$ consecutive bases are connected with an edge. \textbf{Contigs} are contiguous sequences of $k$-mers (i.e.\ $k$-mers that are error-free with high confidence) and represent underlying genomic regions. Contigs are typically longer than the input reads. Finally,  \textbf{scaffolds} are long genomic sequences that consist of oriented contigs which are stitched together. 

The genomes comprising a metagenome dataset have generally variable read coverage, since some
species may exist in the environmental sample with much higher abundance than others. Choosing an
optimal value of $k$ for the de Bruijn graph is therefore challenging because there is a tradeoff in
$k$-mer size that affects high and low frequency species differently. Typically, a small $k$ is
appropriate for low-coverage genomes since it allows a sufficient number of overlapping $k$-mers to
be found and as a result the underlying sequences can be assembled to contigs. On the other hand, a
large $k$ is better suited for the high-coverage genomes since a sufficient number of
overlapping, long $k$-mers can be found and repetitive regions are disambiguated by such long
$k$-mers.

Iterative contig generation (Algorithm~\ref{alg:cont} and Figure~\ref{fig:pipeline}) aims to eliminate the quality trade-off that
different $k$-mer sizes induce in de Bruijn graph-based
assemblers~\cite{peng2012idba,li2015megahit,nurk2016metaspades}. The algorithm starts by
constructing the de Bruijn graph with a small $k$ and extracts a set of contigs by traversing
the graph. After performing a series of transformations on the set
of contigs, $k$ is increased by a step size $s$ and MetaHipMer builds the corresponding de Bruijn graph
from the input reads with ($k+s$)-mers while the graph is enhanced with ($k+s$)-mers extracted from
the previous contig set. This iterative process is repeated until $k$ reaches a user-specified
maximum value.

The quality of the assembly is improved by additional transformations that refine the de Bruijn
graphs (as shown in steps 3 to 6 in Figure~\ref{fig:pipeline}). More specifically, ``bubble
structures" are merged and ``hair" tips (short, dead-end dangling contigs) are removed since they
are potentially created from erroneous vertices. Then, the graph is iteratively pruned in order to
eliminate branches that do not agree with the coverage of the neighboring vertices; such branches
are likely to be created by erroneous edges. Finally, a local assembly algorithm extends
the contigs remaining in the de Bruijn graph, using localized reads aligned to each contig, enabling
the retrieval of $k$-mers which otherwise would be
excluded from the de Bruijn graph because of global conflicts.

Before describing the stages of the MetaHipMer pipeline, we provide an overview of distributed hash tables in UPC; this data structure is the backbone of all of our parallel algorithms.
\subsection{High Performance Distributed Hash Tables in MetaHipMer}
\label{subsec:hash}
Our hash tables utilize a chaining rule to resolve collisions in the buckets. The hash table entries are stored in the shared address space of UPC and thus they can be accessed by any processor with simple assignment statements. This feature of UPC facilitates the design of highly irregular, distributed memory algorithms via a shared-memory programming paradigm. Note that the hash tables involved in our algorithms can be gigantic (hundreds of Gbytes up to tens of Tbytes) and cannot fit in a typical shared-memory node. Therefore it is crucial to distribute the hash table buckets over multiple nodes and in this quest the global address space of UPC is convenient. Here we identify a handful of use cases for the distributed hash tables that allow specific optimizations in their implementation. These use-cases will be used as points of reference in the sections that detail our parallel algorithms.
\begin{itemize}
\item \textbf{Use case 1 -- Global Update-Only phase}\\
\end{itemize}\vspace*{-\baselineskip}
The operations performed in the distributed hash table are only global updates with commutative
  properties (e.g. inserts only). The global hash table will have the same state (although possibly
  different underlying representation due to chaining) regardless of insert order. The global
  update-only phase can be optimized by dynamically aggregating fine-grained updates (e.g. inserts)
  into batch updates. In this way we can reduce the number of messages and synchronization
  events. We can also overlap computation/communication or pipeline communication events to further
  hide the communication overhead. An example of this use-case is storing the $k$-mers in a
  distributed hash table in preparation for the de Bruijn graph traversal.
  
\begin{itemize}
\item \textbf{Use case 2 -- Global Reads \& Writes phase}\\
\end{itemize}\vspace*{-\baselineskip}
The operations performed during this phase are global reads and writes over the \emph{already
    inserted entries}. Typically we can't batch reads and/or writes since there might be race
  conditions that affect the control flow of the governing parallel algorithm. However, we can use
  global atomics (e.g. compare-and-swap) instead of fine-grained locking in order to ensure
  atomicity. The global atomics may employ hardware support depending on the platform and the
  corresponding UPC implementation. We can also build synchronization protocols at a higher level
  that do not involve the hash table directly but instead are triggered by the results of the atomic
  operations. Finally, we can implement the delete operation of entries with UPC atomics and avoid
  locking schemes. An example of this use-case is accessing the $k$-mers during de Bruijn graph
  traversal.

\begin{itemize}
\item \textbf{Use case 3 -- Global Read-Only phase}\\
\end{itemize}\vspace*{-\baselineskip}
In such a use case, the entries of the distributed hash table are read-only and a degree of data
  reuse is expected. The optimization that can be readily employed is to design software caching
  schemes to take advantage of data reuse and minimize communication. These caching schemes can
  be viewed as ``on demand" copying of remote parts of the hash table. Note that the read-only phase
  guarantees that we do not need to provision for consistency across the software caches. Such
  caching optimizations can be used in conjunction with locality-aware partitioning to increase the
  effectiveness of the expected data reuse. Initially even if the data is remote, it is likely to be
  reused later locally. An example is the use of software caches for seed lookup during alignment.
\begin{itemize}
\item \textbf{Use case 4 -- Local Reads \& Writes phase}\\
\end{itemize}\vspace*{-\baselineskip}
 In this use case, the entries in the hash table will be further read/written only by the
  processor owning them. The optimization strategy we employ in such a setting is to use a
  deterministic hashing from the sender side and local hash tables on the receiver side. The local
  hash tables ensure that we avoid runtime overheads and also high-performance, serial hash table
  implementations can be seamlessly incorporated into parallel algorithms. For example, consider
  items that are initially scattered throughout the processors and we want to send occurrences of
  the same item to a particular processor for further processing (e.g. consider a ``word-count" type
  of task). Each processor can insert the received items into a local hash table and further
  read/write the local entries from there. An example of this use case is the distributed histogram
  that gets constructed during $k$-mer analysis.

We emphasize that this is not an exhaustive list of use cases for distributed hash tables. Nevertheless, it captures the majority of the computational patterns we identified in our parallel algorithms that will be detailed in the following sections. In the following subsections we describe the various stages of iterative contig generation.
\subsection{K-mer Analysis using Distributed Histograms}
\label{sec:kmer-analysis}

The first step of the contig generation is parallel $k$-mer analysis, which splits the input
reads into $k$-mers that overlap by $k-1$ consecutive bases, keeping a count for each $k$-mer
occurring more than $\epsilon$ times ($\epsilon \approx 2, 3$) in order to implicitly exclude
sequencing errors. $K$-mer analysis additionally requires keeping track of all possible
\emph{extensions} of each $k$-mer from either side (bases before/after a $k$-mer in a read).
If a nucleotide on an end appears more times than a threshold $t_{hq}$, it is characterized as a
\emph{high quality extension}.

In MetaHipMer, we integrate the parallel implementation of $k$-mer analysis
described in HipMer~\cite{hipmer,merbench}, which uses distributed histograms (Global Update-Only phase and Local Reads \& Writes phase), all-to-all exchanges of
$k$-mers, and distributed Bloom filters (to avoid the memory footprint explosion that is induced by erroneous $k$-mers). Of particular importance to
metagenome assemly, the HipMer implementation uses a specialized streaming algorithm to identify and count ``heavy hitters'', which are $k$-mers that occur millions of times and can potentially cause load imbalance issues if not treated with a specialized algorithm. Such ``heavy hitters" are likely common in metagenomic datasets where highly abundant organisms yield multiple copies of the same $k$-mers.

\subsection{De Bruijn Graph Traversal via a Distributed Hash Table}
\label{sec:dbgraph-traversal}




The de Bruijn graph of the $k$-mers stemming from the $k$-mer analysis is traversed in order to form
\emph{contigs}, which are paths in the de Bruijn graph formed by $k$-mers with \emph{unique} high
quality extensions. These paths represent ``confidently" assembled sequences and can be seen in 
Figure~\ref{fig:pipeline}(a) by removing the branches (vertices with dashed incident
edges) and considering the connected components in the resulting graph. Note that the vertices with
dashed incident edges (i.e.\ ``fork" vertices) do not have unique high quality extensions and can be
used later to discover the connectivity among the contigs. In MetaHipMer, the de Bruijn graph traversal is implemented using a distributed hash table, similar to the approach introduced in HipMer (Global Update-Only phase and Global Reads \& Writes phase). Due to the nature of DNA, the de Bruijn graph is extremely sparse. For example, the human genome's adjacency matrix that represents the de Bruijn graph is a $3\cdot 10^9 \times 3\cdot 10^9 $ matrix with between two and eight non-zeros per row for each of the possible extensions. Using a direct index for the $k$-mers is not practical for realistic values of $k$, since there are $4^k$ different $k$-mers. A compact representation can be leveraged via a hash table: A vertex ($k$-mer) is a key in the hash table and the incident vertices are stored implicitly as a two-letter code [\texttt{ACGT}][\texttt{ACGT}] that indicates the unique bases that immediately precede and follow the $k$-mer in the read dataset. By combining the key and the two-letter code, the neighboring vertices in the graph can be identified. Also, the underlying graph of $k$-mers is characterized by high-diameter where parallel Breadth First Search (BFS) approaches do not scale well and HipMer's specialized traversal overcomes this challenge~\cite{sc14}. However, the HipMer algorithm was designed for single genomes 
and assumes uniform depth coverage. This is usually not the case with metagenomes, where the
coverage of some genomes may be thousands of times higher than others. Thus, the graph traversal
implemented in MetaHipMer differs in this aspect compared to the HipMer implementation.

In HipMer, a $k$-mer with depth greater than $\epsilon$ is
extended in the graph traversal only if there are no more than $t_{hq}$ alternative extensions to
the most common extension of that $k$-mer. The value of $t_{hq}$ is global, and is used for all
$k$-mers. This is potentially a problem in metagenomes because $k$-mers from genomes with
high coverage $COV_{high}$ (e.g.\ abundance in the dataset of 1,000 or more) will typically have
more than $COV_{high}\times e$ alternates if the sequencing error rate is $e$. Therefore, setting a
$t_{hq}$ below $COV_{high}\times e$ can in theory start to fragment high coverage genomes that
are prevalent in the dataset, even though such genomes should be the easiest to assemble. On the
other hand, setting a $t_{hq}$ above $COV_{high}\times e$ will fragment the genomic regions with low
coverage.

The solution we introduce in MetaHipMer is to replace the global threshold, $t_{hq}$, with
one that depends on the depth $d_{k\mbox{-}mer}$ of the $k$-mer that is being extended.  In MetaHipMer,
a $k$-mer with count $d_{k\mbox{-}mer}$ is extended in the de Bruijn graph traversal if
there are no more than $t_{hq} = \max (t_{base}, e \times d_{k\mbox{-}mer})$ extensions that
contradict the most common extension. Here $t_{base}$ is a hard limit in the $t_{hq}$ value and $e$
is a single parameter model for the sequencing error.

\subsection{Parallel Bubble Merging with Speculative Graph Traversal}

%

A single-nucleotide polymorphism (SNP) represents a difference in a base between two genomic
sequences. SNPs create similar contigs (paths in the de Bruijn graph with the same length) except in
one position; these contigs also have the same $k$-mers as extensions of their endpoints and as a result form \emph{bubble} structures in the de Bruin graph~\cite{zerbino2008velvet,hernandez2008novo,simpson2009abyss}. In this step we identify these bubbles and merge them into a single contig. Additionally, dead-end dangling contigs shorter than $2 k$ nucleotides are considered \emph{hair} and are likely to be false positive structures in the graph, hence we remove them~\cite{zerbino2008velvet,simpson2009abyss,li2010novo}. See Figure~\ref{fig:pipeline}(a) for examples of a bubble and a hair contig in the first graph. MetaHipMer also supports optional merging of long bubble-paths (longer than $2k$), similar to the Megahit~\cite{li2015megahit} assembler. This option trades-off contiguity for preserving species/strain variations.

The first step in bubble merging is to build a \emph{bubble-contig} graph, which MetaHipMer does in
parallel by employing a distributed hash table (Global Update-Only phase and Global Reads \& Writes phase). This graph is orders of magnitude smaller than the original $k$-mer de Bruijn graph because the connected components (contigs) of the original graph have been contracted to super-vertices. Once the bubble-contig graph is built, it is traversed to merge eligible contigs (e.g. by picking one of the contigs from the bubble structures). This parallel traversal uses a speculative algorithm. The processors pick random seeds (contigs) from the bubble-contig graph and initiate independent traversals. Once an independent traversal is terminated, we store the resulting path. However, if multiple processors work on the same path, they abort their traversals and allow a single processor to complete them. More specifically, each vertex (contig) has a ``used" binary flag that indicates if this vertex has been traversed and the processors atomically set this flag for the vertices they are visiting (Global Reads \& Writes phase of hash table). If a processor attempts to traverse a ``used" vertex/contig, then it infers that yet another processor works on the same path and aborts the current traversal. Eventually, processor 0 picks up the aborted traversals and completes them.

\subsection{Iterative Graph Pruning}

\begin{algorithm}[!t]
\begin{algorithmic}[1]
\State{\textbf{Input:} A contig set $C$, \text{length} $k$ \text{and thresholds} $\alpha$, $\beta$, $\tau$}
\State{\textbf{Output:} A pruned contig set $C_{pruned}$}
\State $\tau \gets 1$
\While{$\tau <$ maximum contig depth of $C$}\For {each contig $c \in C$}
\If {length($c$) $\leq 2\cdot k$ {\textbf{and}} \\ \ \ \ \ \ \ \ \ \  depth($c$) $\leq$ min($\tau$, $\beta \  \cdot \ $neighbors-depth($c$))}
\State{Remove $c$ from $C$}
\EndIf
\EndFor
\State $\tau \gets \tau \cdot (1+\alpha)$
\EndWhile
\State $C_{pruned} \gets C$
\end{algorithmic}
\caption{Iterative graph pruning}
\label{alg:pruning}
\end{algorithm}

The remaining graph after bubble merging and hair removal is iteratively pruned in order to eliminate branches that do not agree with the coverage of the neighboring vertices. Such branches are likely to be created by false-positive edges in the contig graph due to sequencing errors. Algorithm~\ref{alg:pruning} implements an iterative pruning strategy similar to the pruning module in IDBA-UD~\cite{peng2012idba}.
%

The parallel version of Algorithm~\ref{alg:pruning} starts by reading in parallel the contig set $C$
along with their depths and also stores the $k$-mers from the $k$-mer analysis step in a distributed
hash table (Global Update-Only phase). In particular, we are interested in the ``fork" $k$-mers since they contain information
regarding the connectivity of the contigs; in graph (a) of Figure~\ref{fig:pipeline} the vertices
with dashed incident edges represent ``fork" $k$-mers. Each one of the $P$ processors is then
assigned $1/P$ contigs; the processor extracts the last $k$-mers in the two endpoints of each contig $c$, looks them up in the distributed hash table and gets the contig-neighborhood information for $c$. The parallel execution then proceeds in the main loop (line 5) of Algorithm~\ref{alg:pruning}. Each
processor visits the contigs assigned to it, and if a contig is both short and has relatively small
depth compared to its neighborhood (lines 6 and 7), it is removed from the contig graph. At the end
of the iteration, each processor updates the neighborhoods of its contigs since some may have been
removed. The depth-cutoff threshold $\tau$ is then increased geometrically and the algorithm proceeds to the next iteration.

The parallel algorithm terminates if no contigs are pruned by any processor during an iteration. In
order to detect if any contigs have been pruned from the graph: (1) every processor sets a local
binary variable \texttt{pruned\_flag} to 1 if \emph{any} of its contigs have been pruned, otherwise
the binary variable is set to 0, and (2) we perform an \texttt{all-reduce} operation on the
\texttt{pruned\_flag} variables with the \texttt{max} function as argument. If the max-reduction
result is 0, no changes have been made in the contig-graph and the parallel algorithm terminates
(i.e.\ it has reached a converged state).

\subsection{Alignment of Reads to Contigs}

In this step of the pipeline the goal is to map the original reads onto the pruned contigs. This
mapping provides information about the read pairs that are aligned towards the ends of the contigs
(e.g.\ Figure~\ref{fig:pipeline}(d)). We determine this mapping using merAligner~\cite{meraligner}, a distributed memory, scalable, end-to-end parallel sequence aligner that implements a seed-and-extend algorithm. 


\subsection{Local Assembly with Dynamic Work Stealing}

In this step we try to extend the remaining contigs using a local assembly methodology that
leverages the alignments of reads to contigs. Because the assembly is localized, erroneous $k$-mers
stemming from high-coverage regions are isolated from similar $k$-mers in low-depth areas, so we can
retrieve $k$-mers which otherwise would be excluded from the de Bruijn
graph. Figure~\ref{fig:pipeline}(e) shows that after local assembly, the contigs have been extended with the orange vertices via ``mer-walks".

For each contig, we first accumulate all reads that can be used to extend that contig. Each thread
reads a portion of the reads file, and stores the reads into a global hash table. Then each thread
processes a local subset of contigs, and extracts the reads relevant to each contig to local storage. The
reads selected are those that can be aligned onto a contig and whose paired reads do not align onto the same
contig. In addition, for paired reads we can use the library insert size to project unaligned reads
to either side of a contig (e.g.\ see Figure~\ref{fig:pipeline}(e)).  Second, the reads are used to
extend the contigs through \emph{mer-walking}, which is a modified, localized version of the contig
generation extension algorithm described earlier in Section~\ref{sec:dbgraph-traversal}. The first
modification is that extension bases are accepted or rejected based on the number and quality
category of the extending bases, which allows for uncontested extensions of lower quality than used
in the original $k$-mer analysis. The second modification is that the mer-size used is dynamically
adjusted in an iterative loop, being upshifted (increased by $L$) when a fork is encountered, or
downshifted (decreased by $L$) when no extensions are encountered (a deadend). The walk terminates
when it encounters a fork after downshifting, or a deadend after upshifting.

Once the involved reads are localized, the actual mer-walking of the gaps does not require
communication and is embarrassingly parallel (Local Reads \& Writes phase). However, if the contigs are statically assigned to
processors, severe load imbalance can occur because the computational cost of walking the contigs
exhibits a high degree of unpredictable variability. To ameliorate this problem, we implement a simple dynamic
work-stealing strategy. Each processor performs a block of independent walks, and upon completion, 
uses a global atomic variable to select another block without overlap with
other processors. Although we use a single global atomic, in practice, we achieve good load balance
with large block sizes, resulting in few steals and little contention.

\subsection{Merging $k$-mer Sets via a Distributed Hash Table}

The MetaHipMer pipeline utilizes the final contigs $C$ of iteration $i$ to enrich the $k$-mer set that will be generated by $k$-mer analysis in the following iteration $i+1$. Since contigs in $C$ were assembled with a smaller $k$ value than the one that will be used in the $i+1$ iteration, $k$-mers stemming from low coverage organisms are likely represented in $C$, while such organisms may not be represented via confident, ``error-free" $(k+s)$-mers in the result of the following $k$-mer analysis. Therefore we extract from contigs in $C$ all the $(k+s)$-mers (henceforth called $prev\_k\mbox{-}mer\_set$) and treat them as ``error-free" $(k+s)$-mers with unique high quality extensions in the $i+1$ iteration (see arrow 7 in Figure~\ref{fig:pipeline} where the orange $(k+s)$-mers are extracted from the final contigs of the previous iteration).

The  $prev\_k\mbox{-}mer\_set$ has to be merged with the ($k+s$)-mers stemming from the original reads and are generated by the $k$-mer analysis of the $i+1$ iteration. First, all processors store in parallel the new $k$-mers resulting from the $k$-mer analysis step in a distributed hash table (Global Update-Only phase). Then, they extract in parallel $(k+s)$-mers from contigs in $C$ and store them in the same distributed hash table. The resulting distributed hash table represents the merged $k$-mer set, where duplicated $k$-mers (existing in both $prev\_k\mbox{-}mer\_set$ and the new $k$-mer set from $k$-mer analysis) are collapsed in a single occurrence. All the $k$-mer stores in the distributed hash table are done via aggregated, asynchronous one-sided messages. 

\subsection{Minimizing Communication via Read Localization}\label{sec:localization}
It has been shown~\cite{meraligner,hipmer,europar} that the reads to contigs alignment step is dominated by fine-grained, irregular lookups of seeds (substrings of reads) in a distributed seed index (hash table that indexes the contigs). The authors of merAligner have therefore implemented a software cache to exploit potential seed index reuse and avoid off-node communication in a distributed memory environment. However they also point out that the input read files do not exhibit any inherent locality, hence at large scale the expected data reuse (and consequently the software cache benefit) is provably limited.


However, MetaHipMer uses iterative contig generation, which presents an opportunity to infer read locality
in the first iteration to improve performance in subsequent iterations. Reads that align onto the
same contig region should be similar, and hence most of the substrings of every read (the seeds)
should be identical to substrings of other reads aligned to the same contig region. If these reads
are assigned for alignment to the same processor, merAligner's software cache will be able to serve
most of the pertaining seed lookups, reducing significantly the off-node communication and the total
execution time (Global Read-Only phase).

Being motivated by the aforementioned observations, we implement a parallel read localization
algorithm to speedup the alignment steps in the following iterations. Given the first set of reads
to contigs alignments, each processor assesses in parallel an equal chunk of the
alignments. Assuming a read $R$ is aligned to contig $c_R$, the processor sends $R$ to the processor
with id $(c_R\ \text{mod}\ P)$, where $P$ is the number of available processors. We leverage the
one-sided communication capabilities of UPC and all the reads are distributed via aggregated,
asynchronous messages. As a result of these read redistributions, all the reads that are mapped to
a contig $c_R$ (which, according to the previous reasoning, are similar) will be sent to the same
processor. In the subsequent iterations of the pipeline, this shuffled set of reads is used for
enhanced locality and to minimize communication incurred in the alignment steps.

An additional side-effect of read localization also benefits the $k$-mer analysis phase. In the $k$-mer analysis phase, when a processor receives a bunch of $k$-mers from remote processors, it updates a locally owned hash table that keeps the individual counts of the received $k$-mers. In principle, this local hash table update is characterized by low locality, since the received $k$-mers are uniformly spread out based on a hash function; hence the memory accesses pertaining to the hash table update exhibit little to no cache reuse. However, after read localization we expect (with high probability) almost all the occurrences of the same $k$-mer to be sent by a remote processor \emph{in the same aggregated message}. As such, when the receiving processor tries to insert the $k$-mers in the local counting hash table, most of the updates will result in cache hits and will improve the attained performance.

\section{Scaffolding}

\begin{algorithm}[!t]
\begin{algorithmic}[1]
\State{\textbf{Input:} A set of paired reads $R$ and a set of contigs $C$ }
\State{\textbf{Output:} A set of scaffolds $S$}
\State $Alignments \gets \ $\Call{AlignReadsToContigs}{$R,C$}
\State $Links \gets \ $\Call{GenerateLinks}{$C, Alignments$}
\State $\mathit{Gapped\_scaffs} \gets \ $\Call{ContigGraphTraversal}{$C,Links$}
\State $\mathit{S} \gets \ $\Call{GapClosing}{$\mathit{Gapped\_scaffs},R,C, Alignments$}
\State $\Call{Return}{}\ S$
\end{algorithmic}
\caption{Scaffolding}
\label{alg:scaffolding}
\end{algorithm}

\begin{figure}[t!]
\centering
\includegraphics[width=\columnwidth]{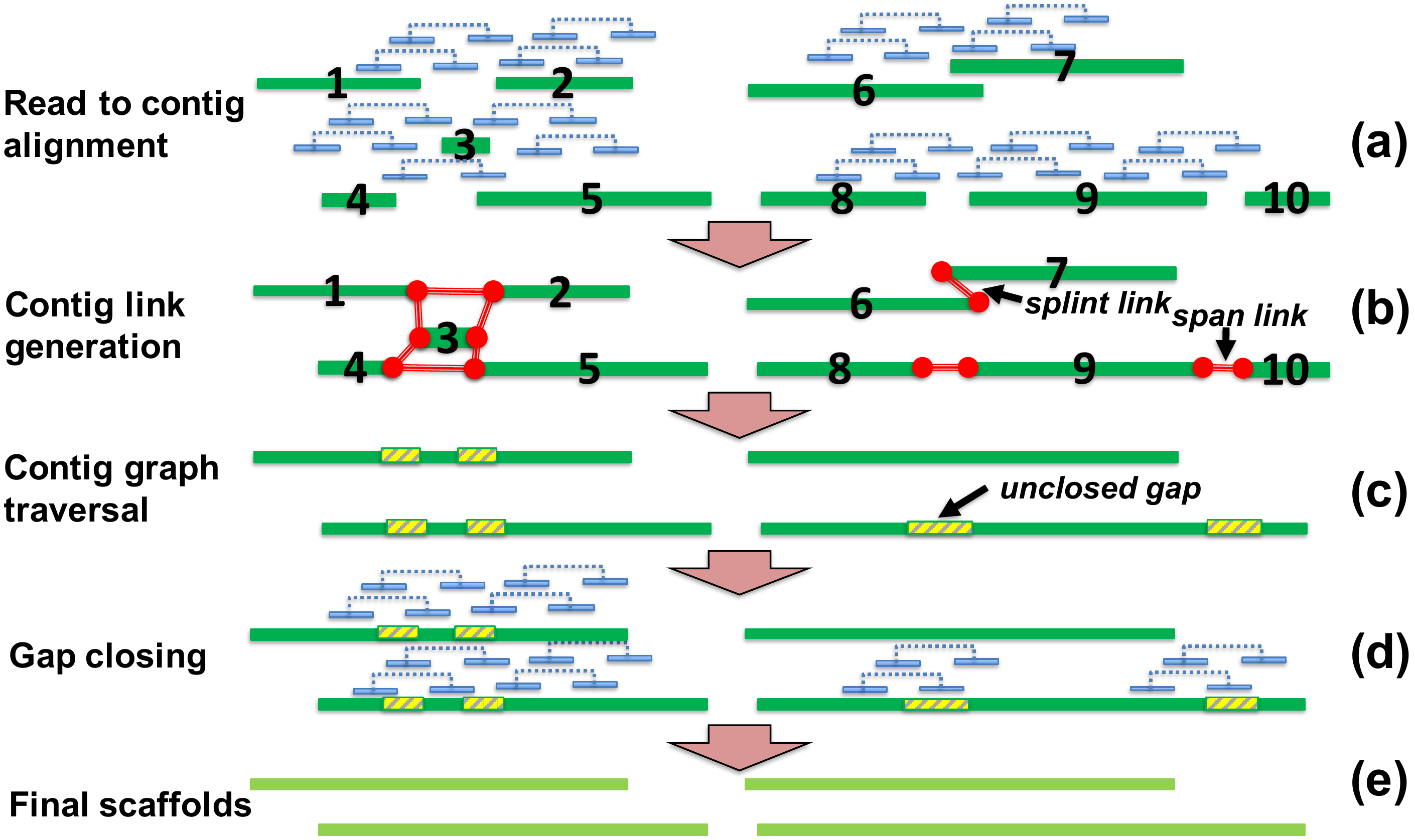}
\caption{Scaffolding workflow in MetaHipMer.}
\vspace{-15pt}
\label{fig:scaff}
\end{figure}

\label{sec:scaffolding}
The main goal of the scaffolding Algorithm~\ref{alg:scaffolding} in MetaHipMer is to connect together contigs and form \emph{scaffolds} which are long chains of contigs. The first step of scaffolding comprises of aligning the input reads onto the contigs generated by the iterative algorithm. Then, by leveraging the reads to contigs alignments and the information from paired reads we introduce additional links/edges in the graph of contigs which we call henceforth \emph{contig graph}. Note that paired reads with large insert sizes can be used to generate long-range links among contigs that could not be found from the $k$-mer de Bruijn graph. Afterwards, we traverse the updated contig graph and form chains of contigs that constitute the final scaffolds. In the following subsections we give more details regarding the scaffolding submodules.

\subsection{Alignment of Reads onto Contigs}
In this step of the pipeline the goal is to map the original reads onto the final contigs generated from the iterative contig generation. This mapping provides information about the relative ordering and orientation of the contigs. Again here we employ merAligner.

\subsection{Contig Link Generation with Distributed Hash Tables}
The next step is to process the alignments and identify \emph{splints}, which are single reads that bridge the gap between two neighboring contigs by virtue of aligning to both of them. Essentially, if a particular segment of a read aligns to the ends of two different contigs we conclude that these contigs form a splint (see Figure~\ref{fig:scaff}(b) for a splint example between contigs 6 and 7). Additionally, by processing paired reads's alignments we identify \emph{spans}, which are read pairs associated with particular pairs of contigs. For example, consider that the first read of a pair aligns with contig $i$ while the second read of that pair aligns with contig $j$. It can thus be concluded that the read pair forms a span (see Figure~\ref{fig:scaff}(b) for a span example between contigs 9 and 10). Also, we know the insert size of the read library and therefore we can estimate the gap size between contigs $i$ and $j$. Once splints and spans are created, they can be aggregated to generate \emph{links} among pairs of contigs. More specifically, if a sufficient number of splints supports a particular distance and mutual orientation between contig $k$ and contig $m$, we generate a \emph{SPLINT link} for that pair of contigs. Analogously, if a sufficient number of paired reads's alignments supports a particular span between contig $i$ and contig $j$ we generate a \emph{SPAN link} for that pair of contigs. 

Regarding the parallelization of the SPLINT-link generation, first each of the $P$ processors independently  processes $1/P$ of the total read alignments and stores the splints's information locally. Then, a distributed hash table is required, where the keys are pairs of contigs and values are the splint/overlap information. Each processor is accessing the local splints and stores them in the distributed hash table. Here, we again apply aggregated, one-sided asynchronous messages to minimize the number of messages and the synchronization cost (Global Update-Only phase). When all splints are stored in the distributed hash table, each processor iterates over its local buckets to further assess/count the splint entries (Local Reads \& Writes phase). The parallel algorithm for the SPAN-link generation is identical to the one for SPLINT-links.

\subsection{Contig Traversal with Connected Components Partitioning}
The splint and span links from the previous step provide essentially the edges in the contig graph (see Figure~\ref{fig:scaff}(b) for a contig graph where the vertices are the contigs --- green pieces -- and the edges are the splint/span links --- red pieces). By traversing this contig graph we form sequences of contigs we call \emph{scaffolds}. The traversal is done by selecting traversal seeds (traversal seeds are contigs) in order of decreasing length; this heuristic tries to ``lock" together first long, confident contigs (the classification into long and short contigs relies on a user-defined threshold). 

There are numerous heuristics involved in the traversal of the contig graph. We call a contig's end \emph{extendable} if it does not have any competing links (links to multiple contigs's ends projected in similar distance from that end). First, edges between long contigs and extendable ends are prioritized in the traversal. If no such edge exists, then we traverse the edge pointing to the closest extendable contig's end; we estimate the distance between contigs's ends based on the links's gap size information.

The contig graph traversal also attempts to resolve repeats. Repeat contigs are typically linked to multiple contigs on both of the endpoints as shown in Figure~\ref{fig:scaff}(b), where contig 3 is a repeat contig and is connected to four contigs 1, 2, 4 and 5. Repeat contigs create competing links and hinder further traversal of the graph. However, if there are span links that unambiguously ``jump over" a repeat contig and connect distant pairs of contigs, then the repeat contig is suspended from the graph, effectively removing competing links and allowing further extensions. For instance, contigs 1 and 2 have a span link that jumps over the repeat contig 3 and as such the latter can be suspended and the repeat can be resolved. The contigs that are classified as suspendable should have length at most equal to the insert-size of the library under consideration. Finally, the suspended contigs will be reconstructed during the gap closing module described in the next subsection.

Another metagenome specific rule we introduce in MetaHipMer's contig graph traversal involves contigs that belong in conserved ribosomal genomic regions. Accurate and effective reconstruction of such ribosomal regions is important for downstream metagenome analysis, e.g.\ for reconstructing phylogenies~\cite{rna}. Therefore, MetaHipMer tries to recognize such ribosomal contigs by using profile Hidden Markov Models (HMM) and in particular we integrate the HMMER pipeline~\cite{wheeler2013nhmmer}. HMMER builds HMM models of these ribosomal regions and efficiently identifies if a given contig fits the HMM models; in this case we call such a contig an \emph{HMM hit}. If a contig of sufficient length is recognized as HMM hit, then we designate both of its ends as \emph{extendable} even in presence of competing links. With an HMM hit contig as source, we initiate aggressive depth first search traversal and we aim to build paths that contain other contigs with similar average $k$-mer depths, which are also HMM hits. This approach allows us to reconstruct long pieces of conserved ribosomal regions without sacrificing accuracy.

The parallelization of the contig graph traversal is non-trivial due to multiple reasons. First, the traversal is done by selecting traversal seeds in order of decreasing length and this rule is fundamentally sequential. Second, the metagenomic-specific rule described in the previous paragraph relies on depth-first search, which is known to be difficult to parallelize. We overcome these parallelization roadblocks by exploiting the nature of the metagenome's contig graph. In particular, we observe that contigs should form connected components/clusters in the contig graph. These clusters can be processed in parallel and we can thus apply our contig graph traversal algorithm independently on each cluster. 

For instance, in Figure~\ref{fig:scaff}(b) we see that there are three independent clusters of contigs. The first step in order to extract parallelism is to identify the connected components in the contig graph. We implemented a simple variant of the Shiloach-Vishkin~\cite{SV} algorithm which is trivially parallelized. We further increase the efficiency of our approach by excluding from the contig graph links with multiplicity less than a user specified threshold. We know that such links will be rejected during the graph traversal algorithm as they are considered unreliable. By excluding such links, we decrease the connectivity of the contig graph and we extract more connected components, or equivalently expose more parallelism for the contig graph traversal. After discovering the connected components, we randomly assign them to processors in order to minimize load imbalance. Finally each processor concurrently traverses the assigned connected components to form scaffolds.
 
\subsection{Gap Closing with Load Balancing}
After the scaffold creation it is possible that there are gaps between pairs of contigs. Figure~\ref{fig:scaff}(c) shows an example where three out of four generated scaffolds contain unclosed gaps. Therefore, we further process the reads to contigs alignments and locate the reads that are placed into these gaps. In MetaHipMer we adopt the parallel gap closing algorithm of HipMer~\cite{hipmer} which has been shown to scale efficiently. The alignments are processed in parallel and projected into the gaps (Global Update-Only phase of hash tables). These  gaps are then divided into subsets and each set is processed by a separate processor, in a completely parallel phase.



Several methods are available for constructing gap closures~\cite{hipmer} and they differ substantially in computational intensity. Given that it is not predicable \emph{a priori} which method will  successfully close a gap, the computational time can vary by orders of magnitude from one closure to the next. To prevent load imbalance in the gap closing phase, the gaps are distributed in a Round Robin fashion across all the available processors. This suffices to prevent most imbalance because it breaks up the gaps from a single scaffold, which tend to require similar costs to close. The outcome of this step constitutes the result of the MetaHipMer assembly pipeline, which are gap closed scaffolds (see Figure~\ref{fig:scaff}(e)).

\section{Results}
\label{sec:results}

This section presents experimental results that demonstrate MetaHipMer's
efficient scalability to thousands of cores on a distributed memory supercomputer, while producing
results comparable in quality to state-of-the-art metagenome assemblers.

\subsection{Experimental Datasets}

\paragraph*{MG64} This is a synthetic dataset comprising a mixture of 64 diverse bacteria and
arterial microorganisms\cite{shakya}. It totals 108.7 million
paired-end Illumina HiSeq 100-pb reads, for a total size of 24GB.

\paragraph*{Wetlands} This is massive-scale metagenomics dataset, containing wetlands soil samples
that are a time-series across several physical sites from the Twitchell Wetlands in the San
Francisco Bay-Delta~\cite{SRA,wetlandsSRAs}. It totals 7.5 billion paired-end Illumina HiSeq reads in 21
lanes, for a total size of 2.6 TBytes. To the best of our knowledge this is the largest metagenomic soil
sample ever collected.

\paragraph*{MGSim}
To conduct a weak scaling performance analysis of MetaHipMer, we developed a tool for generating arbitrarily large and
complex metagenome assembly inputs, called MGSim. MGSim samples
multiple genomes and utilizes the short-read simulator WGSim~\cite{li2011wgsim} to generate reads. The
genomes are sampled with weights calculated from a phylogenetic tree,
and each sampled genome is assigned a relative abundance drawn form a
log-normal distribution.

The BB tools (https://sourceforge.net/projects/bbmap/) with default parameters were used for adapter trimming and removing typical contaminants from all the datasets. For the evaluation we used metaQUAST 4.3\cite{metaquast} with default parameters. 

\subsection{Quality Assessment}
{
\setlength{\tabcolsep}{2pt}
\begin{table}[t!]
\centering
\footnotesize
\begin{tabular}{ |c | c | c | c | c | c | c |c|}
\hline
\textbf{Assembler}  & \multicolumn{3}{c|}{\textbf{Length mbp}
  $\uparrow$} & \textbf{MSA} $\downarrow$ &\textbf{rRNA}  $\uparrow$&
\textbf{Gen.} $\uparrow$ & \textbf{Runtime} $\downarrow$\\
                    & \textbf{$>$5k} & \textbf{$>$25k} &
                    \textbf{$>$50k} &        & \textbf{count}  &
                    \textbf{frac.}  & \textbf {(minutes)} \\ \hline 
\textbf{MetaHipMer}        & 167         &    130         & 108            &
682    & 79             & 94 & 42 \\
\textbf{MetaSPAdes} & 180         & 140            & 115            &
914    & 50             & 94& 73  \\
\textbf{Megahit}    & 177         & 132            & 103            &
761    & 68             & 95 & 21  \\
\textbf{Ray Meta}   & 146         & 106            & 79             &
793    & 75             & 88 & 107 \\
\textbf{HipMer}     & 134         & 74             & 39             &
242    & 28             & 85 & 15 \\ \hline 
\end{tabular}
\vspace{2pt}
\caption{\label{tab:resmet}Comparative assembly quality results for MG64. ``MSA'' stands for misassemblies and
  ``Gen. Frac.'' is genome fraction. Higher quality values are better for all metrics except MSA.}
  \vspace*{-0.9cm}
\end{table}}

Quality of assemblies are usually assessed on
datasets with known reference genomes, such as the MG64 dataset.  We therefore conduct quality
experiments using MG64 and compare MetaHipMer to several other metagenome assemblers,
including MetaSPAdes, Megahit and Ray Meta. We also include a non-metagenome assembler, HipMer~\cite{hipmer} (targeted at single genomes) to demonstrate how an assembler without algorithms specifically tailored for metagenomes can underperform on the same dataset. These runs were all carried out on an 80-core Intel\textsuperscript{\textregistered} Xeon\textsuperscript{\textregistered} E7-8870 2.1GHz server, with
500GB of memory. This platform is used because most other assemblers cannot use distributed memory systems and require a large shared-memory node. 

When determining quality, there is a trade-off between contiguity (the length of the assembly),
coverage (how much of the reference was assembled), and correctness. We use several metrics,
determined by running the metaQUAST 4.3\cite{metaquast} with the default parameters. The results are
shown in Table~\ref{tab:resmet}. Contiguity is captured by the \emph{length} metric, which shows how
many base pairs of the assembly are contained in contigs of lengths $\ge5000$, $\ge25000$ and $\ge50000$ base
pairs. As can be seen from the table, MetaHipMer has the second best contiguity, very close to
MetaSPAdes. Coverage is captured by the \emph{genome fraction}, where all the metagenome assemblers
score approximately 94 to 95\%, except for Ray Meta. Broken down into the 64 individual genomes, the
genome fraction is over 80\% for all but one, which is around 4\% (for all assemblers). This latter
genome is very poorly represented in the sample, and so coverage is poor. Finally, correctness is indicated by the
\emph{misassemblies} metric, which shows the number of misassembled scaffolds in the final assembly. As
can be seen from the table, MetaHipMer has the lowest misassemblies count of all the metagenome assemblers
(excluding HipMer). Appendix~\ref{sec:nga50} provides more detailed contiguity/misassembly comparison between MetaHipMer and  MetaSPAdes on MG64.


In Table~\ref{tab:resmet} we also show a metric called rRNA count and it is the number of ribosomal RNA structures found in the assembled genomes. This metric is of particular importance to biologists interested in classifying and identifying the organisms that are being assembled. MetaHipMer finds the most rRNAs, followed closely by
Ray Meta. The quality results for HipMer, the single genome assembler, clearly illustrate why we need
assemblers built specifically for metagenomes. Although HipMer shows low error rates, it does so at
the cost of contiguity (the length over 50k is less than half of MetaHipMer), coverage (85\% compared to
94\% with MetaHipMer) and rRNA (almost 3 times fewer found).

\subsection{Performance Results}

To measure MetaHipMer's parallel scalability, we utilize the NERSC's Cori Cray
XC40 supercomputer, consisting of 2388 compute nodes, each
containing two 16-core Intel\textsuperscript{\textregistered} Xeon\textsuperscript{\textregistered} E5-2698 2.3GHz processors, for a total of 32 cores per node, with
128GB per node. The nodes are connected with a Cray Aries network with Dragonfly topology with
5.625 TB/s total bandwidth. We built the software using Berkeley UPC v 2.26.0 with Intel\textsuperscript{\textregistered} 17.0.2.174 backend compilers.

\begin{figure}[t!]
\centering
\includegraphics[width=0.9\columnwidth]{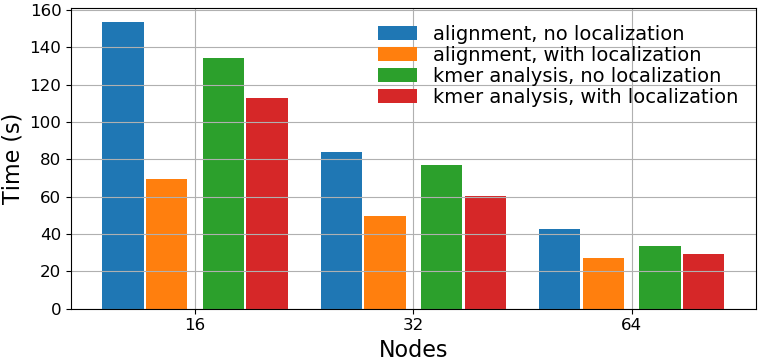}
\vspace{-0.4cm}
\caption{Impact of read localization on $k$-mer analysis and alignment.}
\label{fig:readloc}
\vspace{-13pt}
\end{figure}

\paragraph*{Impact of read localization optimization}
Figure~\ref{fig:readloc} presents the impact of our read localization optimization on two of the pipeline
stages: $k$-mer analysis and alignment, when assembling the MG64 dataset on Cori. The
improvement is especially noticeable at lower concurrencies for alignment, with a 2.2$\times$ speedup
at 16 nodes. In general this optimization improves alignment more than $k$-mer analysis; in regard to the alignment phase, this optimization reduces the off-node communication, while regarding $k$-mer analysis this optimization improves cache reuse on a single node.

\paragraph*{Strong-scaling}

\begin{figure}[t!]
\centering
\includegraphics[width=\columnwidth]{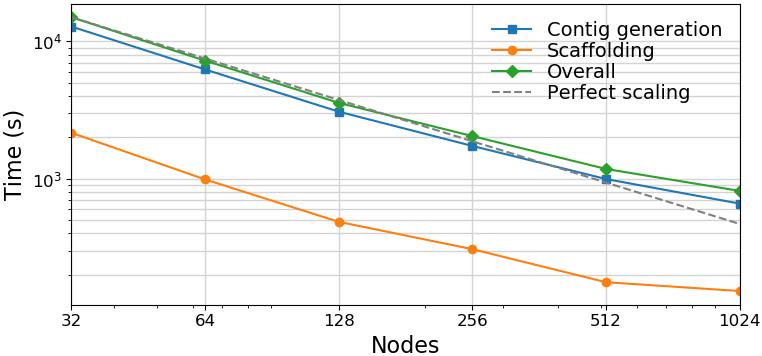}
\vspace{-0.8cm}
\caption{Strong scaling of MetaHipMer on Cori with 3 lanes of Wetlands.}
\label{fig:strongscaling-runtime}
\vspace{-13pt}
\end{figure}

\begin{figure}[t!]
\centering
\includegraphics[width=\columnwidth]{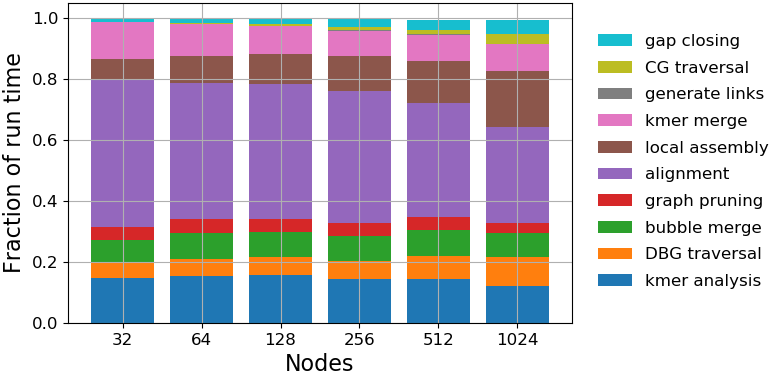}
\vspace{-0.7cm}
\caption{Strong scaling with 3 lanes of Wetlands, runtime fraction of stages.}
\label{fig:strongscaling-stages}
\vspace{-15pt}
\end{figure}

To demonstrate the strong-scaling efficiency, we ran MetaHipMer on a subset of the Wetlands dataset,
consisting of three lanes of reads (about 14\% of the total). To assemble the full dataset requires
at least 512 nodes, and so it is not suitable for strong-scaling
studies. Figure~\ref{fig:strongscaling-runtime} shows strong scaling efficiency of 61\% from 32 (the
minimum required due to memory constraints) to 1024 nodes. The scaling is near perfect until 512
nodes. Most of the computational time is taken by the iterative contig generation phase. The breakdown of MetaHipMer's stages is shown in Figure~\ref{fig:strongscaling-stages}. At smaller
concurrencies, most of the time is taken up by the alignment phase (about 50\%), but at higher
scale, increasing load imbalance in the local assembly stage results
in larger overhead and reduces scalability. As previously
discussed, we implement dynamic work-stealing
for local assembly; this improves load balance from about 0.33 to 0.55 at 1024 cores,
but that is still low enough to cause a gradual drop in overall
scaling. In future work we will address this imbalance by exploiting characteristics of the local
contigs, such as the number of reads that map to a contig.

The only other metagenome assembler (that we are aware of) that scales on distributed memory systems is Ray
Meta~\cite{Ray-Meta}. Ray Meta was too slow to run using the 3-lane Wetlands dataset, so for comparison we use the smaller MG64 dataset. Results show that Ray Meta scales
poorly from 16 to 64 nodes, running at 3,407 secs and 2,931 secs at 16
and 64 nodes respectively (29\% efficiency). By contrast, MetaHipMer takes 512 secs at 16 nodes and 180 secs at 64 nodes (71\% efficiency). At 64 nodes, MetaHipMer is $16\times$ faster than Ray Meta.

\paragraph*{Weak-scaling}

We examine MetaHipMer's weak scaling efficiency by using four datasets generated with MGSim, of increasing size and
complexity. The datasets consist of 5, 10, 20 and 40 genomic taxas that generate 125,
250, 500 million and 1 billion reads, and are run on 128, 256, 512 and 1024 nodes, respectively. 
Table~\ref{tab:weakscale} presents the assembly \emph{rate}, which is defined as kilobases
assembled per second, per node. Results show a slight initial performance
drop fom 128 to 256 nodes, and after that point, the rate remains relatively
unchanged, resulting in 75\% weak scaling efficiency from 128 to 1024 nodes.



\begin{table}[t!]
\centering
\footnotesize
\begin{tabular}{ |c | c | c | c |}
\hline
\textbf{XC40 Nodes} & \textbf{Reads (Million)} & \textbf{Genomic Taxas} & \textbf{KBases/Sec/Node}
\\ \hline
128 & 125 & 5  & 0.16 \\
256 & 250 &10 & 0.12\\
512 & 500 & 20  & 0.13 \\
1024 & 1000 & 40 & 0.12 \\
\hline
\end{tabular}
\caption{\label{tab:weakscale}Metahipmer weak scaling in KBases/Sec/Node on mgsim.}
\vspace*{-0.9cm}
\end{table}

\paragraph*{Grand challenge}
While our strong-scaling results examined performance on a three-lane subset of the Wetlands data,
MetaHipMer enables, for the first time, a full assembly of the 2.6 TByte, 21-lane sample. This took 3
hours and 25 minutes on 512 nodes (16,384 cores) of Cori. Assembling datasets of this size has been
previously proved intractable; we anticipate that this capability will open up a new era in
metagenomic analysis. The benefits of assembling the full dataset over a subset become apparent when
comparing the assembly to that from the three-lane dataset. The full Wetlands assembly is 41.5gbp (giga
base pairs) in length, which is 18$\times$ larger than  the 2.3gbp assembly length for the three-lane subset. Furthermore, the
coverage is much improved, with 42\% of the full set of reads mapping back to the full assembly,
compared to only 7.6\% mapping back to the subset assembly.

\section{Related Work}
As the comparison of the assembly results between metagenome focused assemblers is beyond the scope of this paper, and has been covered in recent work~\cite{CAMI,2017EvalMock,2017CompTools}, we restricted our performance comparison to the Ray Meta~\cite{Ray-Meta}  metagenome assembler that scales on distributed memory systems. Ray Meta is a parallel de novo metagenome assembler based on de Bruijn graphs that utilizes MPI and exhibits strong scaling. One drawback of Ray Meta is the lack of parallel I/O support.  Ray Meta performs best on those organisms that are highly covered within a sample and generally has lower contiguity than MetaHipMer. The results in Section~\ref{sec:results} showed limited Ray Meta parallel efficiency for our test problem. 


MetaSPAdes~\cite{nurk2016metaspades}, is a single-node metagenome assembler that is de Bruijn graph-based, which has excellent quality metrics and performs well on small to medium size datasets.  By default MetaSPAdes includes a read correction stage which we disabled to better match the full workflow of comparable assemblers, including MetaHipMer, when comparing performance, since in most workflows read correction can be treated as pre-processing step before assembly.  MetaSPAdes is limited to problems that fit in the memory of a single node, and thus cannot assemble grand-challenge datasets.

Megahit~\cite{MEGAHIT1.0}, is a single-node metagenome assembler based on de Bruijn graphs that is fast and does an excellent job in assembling low abundance genomes within a metagenome.  Megahit can assemble low and medium sized datasets and is optimized for GPU processing and low memory consumption.

A couple of distributed memory algorithms have been recently developed and tackle only parts of the metagenome assembly pipeline. Kmerind~\cite{pan2017kmerind} is a parallel library for $k$-mer indexing and has been shown to scale efficiently on distributed memory systems. Also, Flick et.\ al.~\cite{flick2015parallel} introduced a parallel connectivity algorithm for de Bruijn graphs in metagenomic applications and the results illustrate very good strong scaling on large concurrencies. However, none of the two aforementioned algorithms constitutes a complete end-to-end pipeline tailored for de novo metagenome assembly.

\label{sec:related}

\section{Conclusions And Future Work}
\label{sec:conclusions}

Metagenomic analysis promises to revolutionize numerous fields of study including biomedicine and environmental sciences.  The de novo assembly of these microbial communities is one of the most challenging problems in bioinformatics due to the computational complexity and irregularity of these huge data sets.  Unlike  state-of-the-art metagenome assemblers that are mostly limited to single-node memory footprints and processing capability, we present MetaHipMer, the first end-to-end, massively scalable, high-quality metagenome assembly pipeline.  MetaHipMer reduces computing runtimes by orders of magnitude and enables a new era of metagenome assemblies that were previously considered intractable.

MetaHipMer's efficient scaling required numerous algorithmic innovations to develop its iterative high-quality approach, coupled with novel parallel computing optimizations. The parallel scalability of MetaHipMer is built on our distributed-memory implementation of irregular data structures, including histograms, hash tables and graphs, which leverage one-sided communication and remote atomics using UPC's global address space capabilities.  Additionally, we employed a variety of techniques to maximize performance including locality-aware hashing, software caching, read localization, partitioning via connected components, and dynamic load balancing.  To evaluate the efficacy of MetaHipMer, we first examined assembly quality metrics across leading metagenome assemblers and demonstrated comparable results on the frequently studied MG64 synthetic data set. We then explored scaling behavior on the Cori Haswell system and showed efficient strong-scaling behavior on up to 1024 nodes (32,768 cores) using a subset (3 lanes) of the Twitchell Wetlands dataset. Next we successfully validated our metagenome assembler's parallel efficiency in a weak scaling regime, by developing MGSim and generating appropriate simulated data sets. Finally, to highlight the new capability of MetaHipMer we conducted a full assembly of the 2.6 TByte Twitchell Wetlands environmental sample --- to the best of our knowledge, the largest, high-quality de novo metagenome assembly completed to date.

Currently, metagenomic studies are conducted overwhelmingly using high-throughput Illumina short read data. As the cost of third-generation sequencing technologies, such as those offered by companies Pacific Biosciences and Oxford Nanopore, continue to come down, metagenomic studies might also benefit from longer reads. However, third-generation longer reads have significantly higher error rates and require more computational power in order to get assembled, which will increase the importance of parallelism in metagenome assemblers. The distributed data structures and techniques presented in this paper are predicted to be instrumental in the large-scale parallel 
assembly of future datasets, regardless of the prevailing sequencing technology.

\bibliographystyle{IEEEtran}
\bibliography{IEEEabrv,main}

\footnotesize{
\noindent Optimization Notice: Software and workloads used in
performance tests may have been optimized for performance only on
Intel microprocessors.  Performance tests, such as SYSmark and
MobileMark, are measured using specific computer systems,
components, software, operations and functions.  Any change to any
of those factors may cause the results to vary.  You should
consult other information and performance tests to assist you in
fully evaluating your contemplated purchases, including the
performance of that product when combined with other products.
For more information go to http://www.intel.com/performance.

\noindent Intel, Xeon, and Intel Xeon Phi are trademarks of Intel Corporation in the U.S. and/or other
countries.
}
\normalsize

\cleardoublepage

\section{Artifact Description Appendix: Extreme Scale De Novo Metagenome Assembly}

\subsection{Abstract}

This artifact description sketches how to compile MetaHipMer on NERSC's Cori system and how the reported performance numbers can be re-measured. The public MetaHipMer release (https://sourceforge.net/projects/hipmer/) includes more information on how to build MetaHipMer on various platforms.

\subsection{Description}

\subsubsection{Check-list (artifact meta information)}

{\small
\begin{itemize}
  \item {\bf Algorithm: } Distributed memory metagenome assembly
  \item {\bf Program: } Available via SourceForge.net: \texttt{https://sourceforge.net/projects/hipmer}
  \item {\bf Compilation: } cmake
  \item {\bf Data sets: } Datasets are publicly available. 
   \item {\bf Run-time environments: } Linux, GasNet, Unified Parallel C, MPI Environment
  \item {\bf Hardware/System: } NERSC's Cori Cray XC40, consisting of 2388 compute nodes, each comprising two 16-core Intel Xeon E5-2698 2.3GHz processors, for a total of 32 cores per node, with 128GB per node. The nodes are connected with a Cray Aries network with Dragonfly topology with 5.625TB/s total bandwidth.
  \item {\bf Execution: } Via shell scripts/job scheduler
  \item {\bf Output: } Detailed timing for each pipeline module is available in log files that are generated automatically
  \item {\bf Experiment workflow: } see below
  \item {\bf Experiment customization: } Different  assembly experiments can be set up via config files, various hardware platforms are supported by the build process.
  \item {\bf Publicly available?: } Yes
\end{itemize}
}

\subsubsection{How software can be obtained (if available)}
Via SourceForge.net 

\subsubsection{Hardware dependencies}
x86 platforms.

\subsubsection{Software dependencies}
\begin{itemize}
  \item Working Message Passing Interface - MPI Environment (Open MPI, MPICH2)
  \item Unified Parallel C - UPC environment (Berkeley UPC $\geq$ 2.20.0)
  \item Working C/C++ compiler (Intel $\geq$ 15.0.1.133, GCC $\geq$ 4.8, CLang $\geq$ 700.1.81 )
\end{itemize}

\subsubsection{Datasets}
All performance runs presented in this work were carried out with real datasets that are publicly available through the Sequence Read Archive \texttt{https://www.ncbi.nlm.nih.gov/sra}. The simulated datasets for the weak scaling experiments were generated by a modification of the wgsim simulator: \texttt{https://github.com/ajtritt/wgsim}

\textbf{MG64} has Sequence Read Archive (SRA) accession \#:SRX200676. 

\textbf{Wetlands} has SRA accession \#:SRR1182407, SRR1184661, SRR403474, SRR404111, SRR404117, SRR404119, SRR404151, SRR404204, SRR407529, SRR407548, SRR407549, SRR410821, SRR437909, SRR5198900, SRR5198901, SRR5198902, SRR5198903, SRR5246785, SRR5246787, SRR5246790, SRR5246791, SRR6203186.

\subsection{Installation}

On NERSC's Cori Cray XC40 system:
\begin{scriptsize}
\begin{verbatim}
HIPMER_ENV_SCRIPT=.cori_deploy/env.sh ./bootstrap_hipmer_env.sh install
\end{verbatim}
\end{scriptsize}
The MetaHipMer distribution has several scripts to support building on multiple platforms including Linux and Mac OS X

\subsection{Experiment workflow}

All datasets and config files are already set up on Cori's scratch filesystem (please contact the authors for the paths on Cori). The distribution provides detailed examples on how to set up the required data sets and config files. Exemplary execution script of MetaHipMer on 32 nodes of Cori:

\begin{scriptsize}
\begin{verbatim}
#!/bin/bash
set -e
export NODES=32
export CORES_PER_NODE=32
export THREADS=$((CORES_PER_NODE*NODES))
export CACHED_IO=1 
export UPC_SHARED_HEAP_SIZE=2000 
export HIPMER_INSTALL=$SCRATCH/hipmer-install-cori/
export PATH=$PATH:$HIPMER_INSTALL

export HIPMER_DATA_DIR=$SCRATCH/hipmer_metagenome_data
export HIPMER_TEST=wetlands_parcc_bbqc

sbatch -N $NODES \
    --ntasks-per-node=$CORES_PER_NODE \
    -q regular \
    -C haswell \
    -L SCRATCH \
    -t 04:00:00 \
    -o ${CONFIG}-${NODES}-%j-cori-haswell.out \
    -J ${CONFIG}-${NODES}.out \
    $HIPMER_INSTALL/bin/test_hipmer.sh
\end{verbatim}
\end{scriptsize}

All other assemblers evaluated in this paper were run with their default/suggested parameters.

\subsection{Evaluation and expected result}

Performance can be simply assessed by log output files that are automatically generated and runtime is reported in seconds. The accuracy of the resulting assemblies are evaluated with metaQUAST 4.3 (\texttt{http://quast.sourceforge.net/metaquast}) -- used default parameters.

\subsection{Experiment customization}

The MetaHipMer distribution supports various exemplary experimental setups with customizable config files, run scripts for various platforms and sample read data sets.


\section{Metahipmer and Metaspades NGA50 comparison on MG64}
\label{sec:nga50}

\begin{figure}[t!]
\centering
\includegraphics[width=\columnwidth]{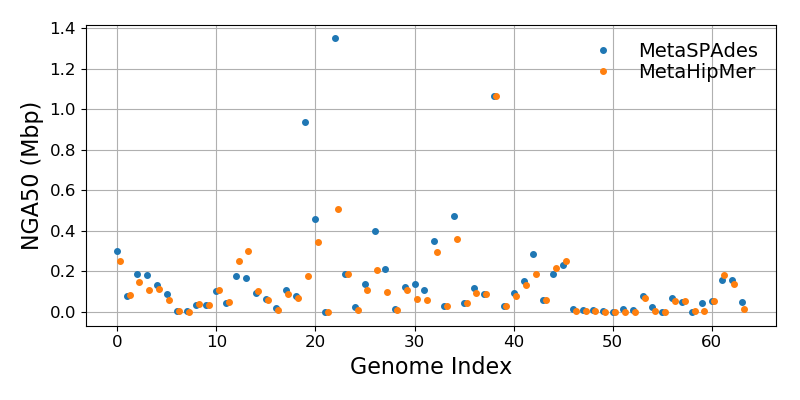}
\caption{All genomes of NGA50 for MG64, comparing MetaHipMer vs MetaSPAdes.}
\label{fig:nga50}
\end{figure}

The results presented in Table~\ref{tab:resmet} show the metrics for the whole MG64 assembly. This
obscures the variation across the 64 genomes that comprise the dataset. To get a better idea of this
variation, Figure~\ref{fig:nga50} presents the NGA50 metric~\cite{metaquast} for each individual
genome. The NGA50 metric is designed to capture contiguity in the presence of errors, and so can be
thought of as a compact measure of both length and misassemblies. We can see from the figure that
MetaHipMer and MetaSPAdes have very similar NGA50 for almost all genomes, except for two outliers. In
these cases, there are so few contigs in the genomes that a single misassembly can change the NGA50 dramatically.

\end{document}